\documentclass[12pt,aps,pra,groupedaddress]{revtex4}
\usepackage{graphicx}
\usepackage{amsmath}
\usepackage{amsfonts}
\usepackage{amssymb}
\usepackage{wasysym}

\newcommand{\omegabold}{\boldsymbol\omega}

\begin{document}

\title{Asymptotic solution for high vorticity regions in incompressible 3D Euler equations}
\author{D.S. Agafontsev$^{1,2}$, E.A. Kuznetsov$^{2,3}$ and A.A. Mailybaev$^{4}$}

\affiliation{\small \textit{ $^{1}$ P. P. Shirshov Institute of Oceanology, 36 Nakhimovsky prosp., 117218 Moscow, Russia\\
$^{2}$ Novosibirsk State University, 2 Pirogova, 630090 Novosibirsk, Russia\\
$^{3}$ P.N. Lebedev Physical Institute, 53 Leninsky Ave., 119991 Moscow, Russia\\
$^{4}$ Instituto Nacional de Matem\'atica Pura e Aplicada -- IMPA, Rio de Janeiro, Brazil}}

\begin{abstract}
Incompressible 3D Euler equations develop high vorticity in very thin pancake-like regions from generic large-scale initial conditions. 
In this work we propose an exact solution of the Euler equations for the asymptotic pancake evolution. 
This solution combines a shear flow aligned with an asymmetric straining flow, and is characterized by a single asymmetry parameter and an arbitrary transversal vorticity profile. 
The analysis is based on detailed comparison with numerical simulations performed using a pseudo-spectral method in anisotropic grids of up to $972\times 2048\times 4096$.
\end{abstract}

\maketitle


\section{Introduction}
\label{Sec:Intro}

The mechanism of vorticity growth in the incompressible 3D Euler equations, in the absence of physical boundary, was addressed in numerous studies because of its relation to a possible finite-time blowup and subsequent transition to turbulence. 
Several analytical blowup and no-blowup criteria were established; see the reviews in~\cite{chae2008incompressible} and \cite{gibbon2008three}. 
In parallel, a large effort was made with numerical analysis.
In one of the early numerical studies~\cite{brachet1992numerical}, Brachet \textit{et al.}  examined evolution of periodic flows in $256^{3}$ grids for random initial conditions and in $864^{3}$ grids for the symmetric Taylor--Green vortex. 
In all cases, the maximum of vorticity was growing nearly exponentially with time, and the regions of high vorticity were confined within pancake-like structures (thin vortex sheets). 
An exact solution of the Euler equations was suggested as a model for asymptotic pancake development, with the relation $\omega_{\max}(t)\propto 1/\ell(t)$ between the vorticity maximum and pancake thickness. 
Since the tendency toward a vortex sheet should suppress three-dimensionality of the flow, formation of a finite-time singularity is not expected; recall that the dynamics of the 2D Euler equations is known to be regular; see~\cite{majda2002vorticity} and the related discussion in~\cite{pumir1990collapsing,ohkitani2008geometrical}. 
Thus, further numerical studies were mainly concentrated on carefully designed initial conditions providing enhanced vorticity growth. 
We refer to~\cite{gibbon2008three,agafontsev2015} for a brief review, as well as to~\cite{hou2009blow,bustamante2012interplay,kerr2013bounds} for examples of recent numerical works. 
It is fair to say that we do not possess a sufficiently reliable evidence supporting the blowup hypothesis yet.

In our previous study~\cite{agafontsev2015}, we returned to the problem of vorticity growth from generic large-scale initial conditions and focused on numerical description of flow details with high resolution. 
Two simulations were carried out with the grids up to $486\times 1024\times 2048$ and $1152\times 384\times 2304$ for initial conditions designated as $I_{1}$ and $I_{2}$. 
Self-similar development of the pancake-like regions of high vorticity was observed. 
However, significantly different exponents ($\beta_{2}/\beta_{1}\approx 2/3$) were measured for the maximum vorticity growth $\omega_{\max}(t)\propto e^{\beta_{2}t}$ and the pancake compression in the transversal direction $\ell(t)\propto e^{-\beta_{1}t}$, demonstrating that the pancake model with $\omega_{\max}(t)\propto 1/\ell(t)$ suggested in~\cite{brachet1992numerical} is insufficient. 
The pancake structures were emerging in increasing number with time. 
These structures provided the leading contribution to the energy spectrum, where we observed the gradual formation of the Kolmogorov spectrum, $E(k)\propto k^{-5/3}$, in a fully inviscid flow. 

In the present paper we demonstrate that the asymptotic pancake evolution can be described by a new \textit{exact} solution of the Euler equations, which combines a shear flow aligned with an asymmetric straining flow. 
This solution represents an essential generalization of the pancake model of~\cite{brachet1992numerical} and agrees with the numerical data. 
We illustrate our results with the simulation of $I_{1}$ initial condition from~\cite{agafontsev2015}, performed here in eight times larger grid up to $972\times 2048\times 4096$, and concentrate our analysis on the main pancake structure containing the global vorticity maximum. 
We checked that other pancake structures, as well as pancakes developing in simulations of different initial conditions, also agree with the exact solution. 

The paper is organized as follows. 
Section~\ref{Sec:NumResults} describes the numerical method and demonstrates general properties of a pancake structure. 
The exact solution of the 3D Euler equations is proposed in Section~\ref{Sec:Solution} as a model for asymptotic pancake evolution. 
In Section~\ref{Sec:Comparison} we provide several numerical tests comparing the analytical model with the simulations. 
The final Section contains conclusions. 


\section{Pancake vorticity structures}
\label{Sec:NumResults}

We analyze evolution of high-vorticity regions with numerical simulations of the Euler equations (in the vorticity formulation)
\begin{equation}
\frac{\partial\omegabold}{\partial t}=\mathrm{rot}\,(\mathbf{v}\times \omegabold),\quad
\mathbf{v} = \mathrm{rot}^{-1}\omegabold,
\label{Euler2}
\end{equation}
in the periodic box $\mathbf{r} = (x,y,z)\in \lbrack -\pi ,\pi ]^{3}$. 
The pseudo-spectral Runge-Kutta fourth-order method is used, together with the Fourier cut-off function suggested in~\cite{hou2007computing} to avoid the bottle-neck instability. 
The inverse of the curl operator and all spatial derivatives are calculated in the Fourier space. 
We start from initial condition $I_{1}$ of~\cite{agafontsev2015}, which represents a perturbation of the shear flow $\omega_{x}=\sin z$, $\omega_{y}=\cos z$, $\omega_{z}=0$.
Taking advantage of the anisotropy of vorticity field, we use an adaptive anisotropic rectangular grid, which is uniform along each direction and adapted independently along each coordinate. 
For more details of the numerical scheme, see~\cite{agafontsev2015}, where it was verified that the accuracy within the simulation time interval is very high and not affected by the Fourier cut-off filter. 
The simulation previously stopped at time $t=6.89$ with the grid $486\times 1024\times 2048$ is continued here until $t=7.75$ with the eight times larger final grid $972\times 2048\times 4096$, when the thinnest pancake structure is resolved with $10$ grid points at the level of vorticity half-maximum.
The results of the two simulations perfectly converge at the same times. 
Both the energy $E=(1/2)\int \mathbf{v}^2\,d^{3}\mathbf{r}$ and helicity $\Omega=\int (\mathbf{v}\cdot\omegabold)\,d^{3}\mathbf{r}$ are conserved with the relative error smaller than $10^{-11}$.

\begin{figure}
\centering
\includegraphics[width=14.0cm]{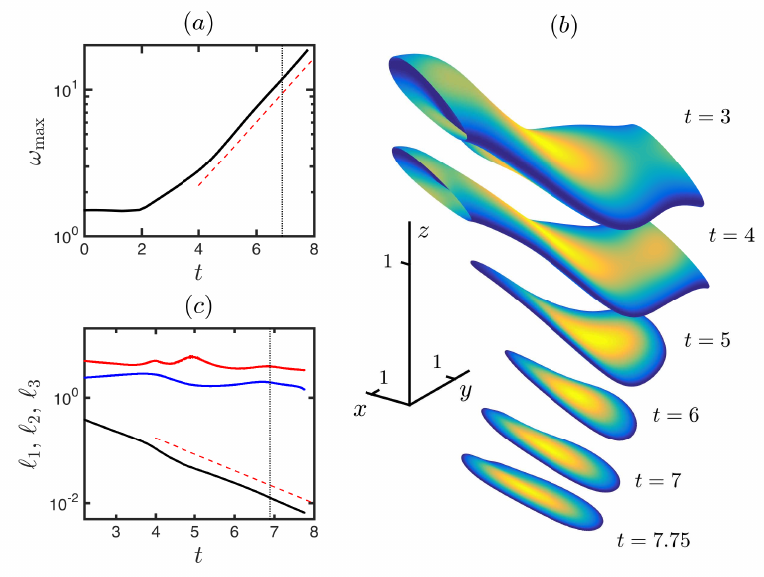}
\caption{
(a) Global vorticity maximum as a function of time (logarithmic vertical scale). The red dashed line indicates the slope $\propto e^{\beta_{2}t}$ with $\beta_{2}=0.5$. The thin vertical line marks the final simulation time  $t = 6.89$ in~\cite{agafontsev2015}. 
(b) Regions of the largest vorticity, $\omega \ge 0.85 \,\omega_{\max}(t)$, at different times. Color represents vorticity in the mid-plane of the pancake: from blue for $0.85 \,\omega_{\max}(t)$ to yellow for $\omega_{\max}(t)$. The structures are shifted vertically for better visualization. 
(c) Characteristic spatial scales $\ell_{1}$ (black), $\ell_{2}$ (blue) and $\ell_{3}$ (red) of the pancake structure. The red dashed red line indicates the slope $\propto e^{-\beta_{1}t}$ with $\beta_{1}=0.74$. 
}
\label{fig:omega_max_dim_main}
\end{figure}

Fig.~\ref{fig:omega_max_dim_main}(a) shows evolution of the global vorticity maximum $\omega_{\max}(t) = \max_{\mathbf{r}} |\omegabold(\mathbf{r},t)|$, demonstrating the vorticity increase from $\omega_{\max}(0) = 1.5$ up to $18.4$ at $t=7.75$ and the asymptotically exponential vorticity growth at late times. 
Panel (b) of the same figure shows the three-dimensional regions containing the vorticity $\omega = |\omegabold| \ge 0.85 \,\omega_{\max}(t)$ at different times, and one can clearly see the formation of a thin pancake structure. It is convenient to introduce the pancake mid-plane as a surface, where the vorticity attains a maximum within the pancake thickness. 
The color in Fig.~\ref{fig:omega_max_dim_main}(b) describes the mid-plane vorticity, from blue for $0.85 \,\omega_{\max}(t)$ to yellow for $\omega_{\max}(t)$. 
At $t = 3$ and $4$, the pancake spans the whole periodic domain in $x$-direction; for larger times its lateral sizes decrease, but eventually stabilize and remain almost constant at $t \ge 6$. 
On the contrary, the thickness keeps decreasing rapidly. 
Thus, at late times, vorticity variations become large (small) in transversal (tangential) directions to the pancake. 

The local geometry of the thin structure can be studied using the Hessian matrix $\partial_{i}\partial_{j}\omega^2$ of second derivatives of $\omega^2$ with respect to $(x,y,z)$, computed at the point of maximum vorticity; the location of the latter in between the grid nodes is approximated with the second-order finite-difference scheme. 
The (unit) normal vector $\mathbf{n}_1$ to the pancake is defined as the eigenvector corresponding to the largest of the three eigenvalues $|\lambda_1|\ge |\lambda_2|\ge |\lambda_3|$ of the Hessian. 
The pancake thickness $\ell_1$, as well as its lateral scales $\ell_2$ and $\ell_3$, can be estimated with the local second-order approximation as $\ell_i = \sqrt{2\omega_{\max}^2/|\lambda_i|}$, see~\cite{agafontsev2015}. 
Fig.~\ref{fig:omega_max_dim_main}(c) shows these characteristic scales as functions of time (the noise in $\ell_{3}$ is due to the ill-conditioned Hessian matrix). 
At late times, the pancake thickness $\ell_1$ is exponentially decreasing, while the dimensions $\ell_2$ and $\ell_3$ do not change substantially, in agreement with the three-dimensional picture in Fig.~\ref{fig:omega_max_dim_main}(b). 
Thus, the pancake develops only one small scale corresponding to its transversal direction, and the vorticity growth is locally one-dimensional. 
The span-to-thickness ratio grows exponentially, reaching $\ell_2/\ell_1\sim\ell_3/\ell_1 \gtrsim 100$ at the final time $t=7.75$. 
The vorticity vector within the pancake structure is tangent to the pancake mid-plane and  oriented roughly anti-parallel to the $y$-axis. 


\section{Exact solution of the Euler equations as a model for pancake evolution}
\label{Sec:Solution}

Following the numerical results, we suggest an analytical model for the vorticity growth. 
Assuming that in Cartesian coordinates $\mathbf{a} = a_{1}\mathbf{n}_{1} + a_{2}\mathbf{n}_{2} + a_{3}\mathbf{n}_{3}$ the vorticity changes only along $a_{1}$-axis and is oriented along $a_{2}$-axis, we write
\begin{equation}
\mathbf{\omegabold}(\mathbf{a},t) = \omega_2\mathbf{n}_2,\quad
\omega_2 = \Omega(t)f'\left(\frac{a_{1}}{\ell_{1}(t)}\right),
\label{eq1_1}
\end{equation}
where $\Omega(t)$ is the characteristic vorticity amplitude and $\ell_{1}(t)$ is the pancake thickness. 
The ansatz (\ref{eq1_1}) contains a derivative (denoted by prime) of an arbitrary function $f(\xi)$ taken at $\xi = a_1/\ell_{1}(t)$. 
One can check that Eq.~(\ref{eq1_1}) together with the velocity field
\begin{equation}
\mathbf{ v}(\mathbf{a},t) = -\Omega(t)\,\ell_{1}(t)\,f\left(\frac{a_1}{\ell_{1}(t)}\right) \mathbf{n}_3
+\left(\begin{array}{r}
-\beta_1(t)\,a_1\\\beta_2(t)\,a_2\\\beta_3(t)\,a_3
\end{array}\right)
\label{eqA2}
\end{equation}
represent an exact solution of the Euler equations (\ref{Euler2}), where $\beta_1(t)$, $\beta_2(t)$ and $\beta_3(t)$ are given by 
\begin{equation}
\beta_1 = -\dot{\ell}_{1}/\ell_{1},\quad 
\beta_2 = \dot{\Omega}/\Omega,\quad
-\beta_1 + \beta_2 + \beta_3 = 0,
\label{eqA3}
\end{equation}
with the dots denoting time-derivatives. 
Velocity~(\ref{eqA2}) is a superposition of the shear flow $-\Omega\,\ell_{1} f(a_1/\ell_{1})\,\mathbf{n}_3$ and the asymmetric irrotational straining flow $(-\beta_1 a_1,\,\beta_2 a_2,\,\beta_3 a_3)$, and satisfies the Euler equations $\dot{\mathbf{v}}+\mathbf{v}\cdot\nabla\mathbf{v} = -\nabla p$ with the pressure 
\begin{equation}
p = 
(\dot{\beta}_1-\beta_1^2)\,\frac{a_1^2}2 
-(\dot{\beta}_2+\beta_2^2)\,\frac{a_2^2}2 
-(\dot{\beta}_3+\beta_3^2)\,\frac{a_3^2}2.
\label{eqA4}
\end{equation}
A uniform velocity field with an arbitrary time-dependency $(0,v_{b2}(t),v_{b3}(t))$ can be added to~(\ref{eqA2}). 
This field describing a drift of the pancake structure leads to the change of pressure as $p\to p - a_{2}(\beta_{2}v_{b2}+\dot{v}_{b2}) - a_{3}(\beta_{3}v_{b3}+\dot{v}_{b3})$.

The suggested solution is an essential generalization of the pancake model of~\cite{brachet1992numerical}; the latter is obtained as a special case with $\Omega=\ell_{1}^{-1}=e^{t/T}$, $\beta_{1,2}=1/T$, $\beta_{3}=0$.
Solution~(\ref{eq1_1})-(\ref{eqA4}) has infinite energy in $\mathbb{R}^{3}$ and allows for an arbitrary time-dependency of $\Omega(t)$ and $\ell_{1}(t)$, in particular, the one leading to a finite-time blowup. 
In addition to an arbitrary function $f(\xi)$, the new solution is characterized by a single dimensionless parameter 
\begin{equation}
\sigma = \frac{\beta_2-\beta_3}{\beta_2+\beta_3}=\frac{2\beta_{2}}{\beta_{1}}-1,
\label{eqA5}
\end{equation}
describing the asymmetry of the straining flow in (\ref{eqA2}).
In our numerical simulations, nearly exponential behavior for $\Omega(t)$ (vorticity maximum) and $\ell_{1}(t)$ is observed, see Fig.~\ref{fig:omega_max_dim_main}(a,c), that corresponds to constant numbers for $\beta_1$, $\beta_2$ and $\beta_3$ in Eq.~(\ref{eqA3}). 
Then, the asymmetry parameter defines the exponent in the power-law relation
\begin{equation}
\Omega(t)\propto \ell_{1}(t)^{-\zeta},\quad 
\zeta = \frac{\beta_{2}}{\beta_{1}} = \frac{\sigma+1}{2},
\label{eqA5b}
\end{equation}
between the vorticity amplitude and pancake thickness. 
Consequently, in the transversal direction the velocity has variation $\delta v_{3}\propto \Omega\,\ell_{1}\propto \ell_{1}^{1-\zeta}$ at the scale of the pancake thickness $\delta a_{1}\sim \ell_{1}$, and this variation vanishes for $\zeta<1$ (i.e., $\sigma<1$) as $\ell_{1}\to 0$.

Solution~(\ref{eq1_1})-(\ref{eqA4}) can be extended for the Navier--Stokes equations with kinematic viscosity $\nu$, if the function $f(\xi,t)$ changes with time as $f_{t}-\frac{\nu}{\ell_{1}^{2}} f_{\xi\xi}=0$. 
The latter becomes the heat equation after a simple transformation of time, $\tau=\int\,dt/\ell_{1}^{2}(t)$. 
For the axisymmetric straining flow with $\beta_{2}=\beta_{3}=\beta_{1}/2$, the suggested solution becomes the special case of the Lundgren stretched-spiral vortex, see~\cite{lundgren1982strained}. 
Note that solutions of the Navier-Stokes equations in the form of stretched vortices embedded in a uniform straining flow were first studied by J.M.~Burgers~\cite{burgers1948mathematical}, see also~\cite{prochazka1998structure,pullin1998vortex,gibbon1999dynamically,maekawa2009existence}.

\section{Comparison with the numerical simulations}
\label{Sec:Comparison}

In simulations, we define the local coordinate system $(a_{1},a_{2},a_{3})$ for the pancake structure in the following way. 
The origin is chosen at the point of the global vorticity maximum, where we also compute the pancake normal vector $\mathbf{n}_{1}$ of the $a_{1}$-axis, as described in Section~\ref{Sec:NumResults}. 
According to the exact solution (\ref{eq1_1}), the $a_2$-axis should be parallel to the vorticity vector $\omegabold$. 
However, in simulations the angle between $\mathbf{n}_1$ and $\omegabold$ may differ from $90^{o}$. 
We checked that this difference, in fact, is tiny, reaching at the final time $0.02^{o}$.
Thereby, we define the $a_2$-axis with the direction $\mathbf{n}_2 = c\, [\omegabold-(\omegabold\cdot\mathbf{n}_1)\mathbf{n}_1]$, where $(\omegabold\cdot\mathbf{n}_1)\mathbf{n}_1$ is a small correction and the prefactor $c$ is chosen such that $\|\mathbf{n}_2\| = 1$. 
Finally, the $a_3$-axis has the direction $\mathbf{n}_3 = \mathbf{n}_1 \times \mathbf{n}_2$. 
Note that this coordinate system is computed at each moment of time, what is necessary to account for a possible drift of the whole structure. 
We choose the vorticity amplitude as the maximum vorticity, $\Omega = \omega_{\max}$, and compute coefficients $\beta_{1}$, $\beta_{2}$ and $\beta_{3}$ according to Eq.~(\ref{eqA3}). 

\begin{figure}
\centering
\includegraphics[width=8.4cm]{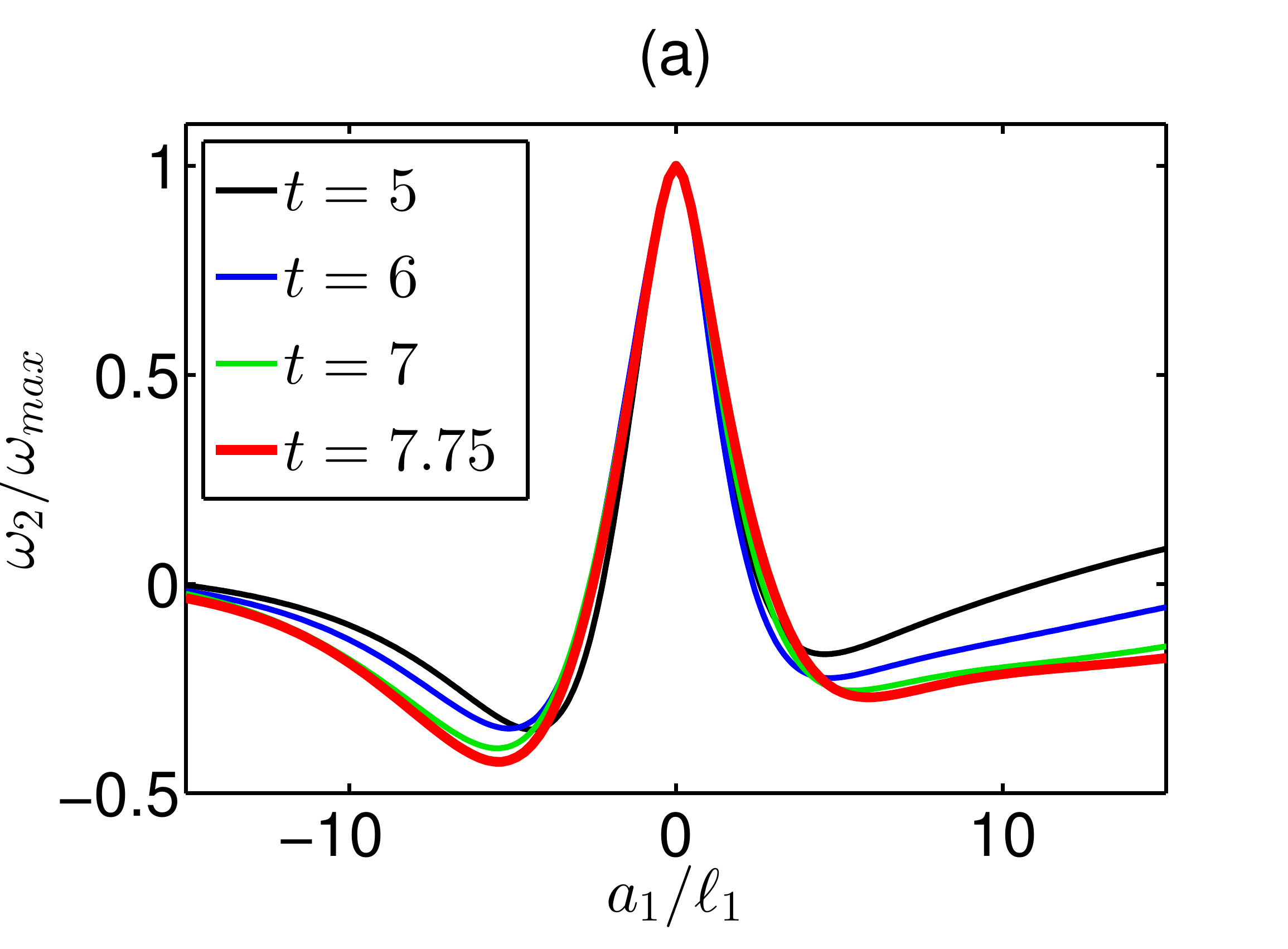} 
\includegraphics[width=8.4cm]{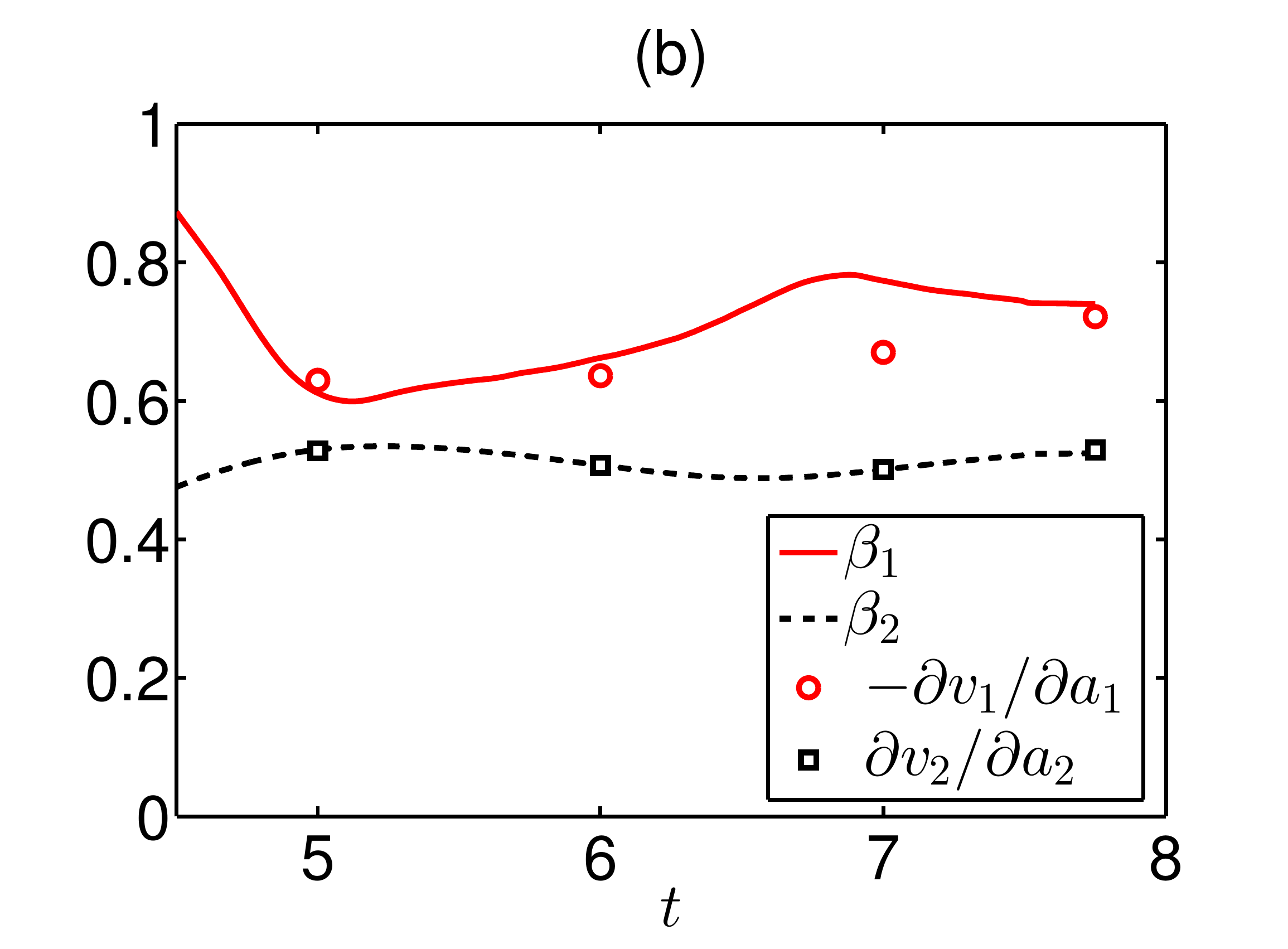}
\caption{
(a) Normalized vorticity component $\omega_2/\omega_{\max}$ as a function of $\xi = a_1/\ell_1$ at different times. 
(b) Comparison of the logarithmic derivatives $\beta_{1}=-\dot{\ell}_1/\ell_1$ and $\beta_{2}=\dot{\omega}_{\max}/\omega_{\max}$ with the velocity gradients $-\partial v_1/\partial a_1$ and $\partial v_2/\partial a_2$ computed at the global vorticity maximum, see Eq.(\ref{eqB2}). 
Prior computing the time derivatives, $\ell_{1}$ and $\omega_{\max}$ are smoothed with the weighted local regression (lowess filter), see~\cite{cleveland1988regression}.
}
\label{fig2}
\end{figure}

In this Section, we provide several numerical tests supporting the pancake model proposed in Section~\ref{Sec:Solution}. 
The first test is related to self-similarity of the transversal vorticity profile, which should be kept according to Eq.~(\ref{eq1_1}), as 
\begin{equation}
\omega_2/\omega_{\max} = f'(\xi),\quad \xi = a_1/\ell_1.
\label{eqB1}
\end{equation}
This is confirmed in Fig.~\ref{fig2}(a), where the vorticity profile $\omega_2/\omega_{\max}$ along the $a_{1}$-axis is shown at different times. 
The self-similarity region grows with time in the $\xi$-coordinate, remaining finite in physical space where it matches with the background flow. 
Note that the function $f(\xi)$ may be arbitrary; in our simulations, different vorticity profiles are obtained for different pancakes and initial conditions. 

For the analysis of the velocity field, one should take into account a drift of the pancake structure by a background flow with nontrivial time-dependency. 
However, this difficulty can be avoided if we examine the velocity gradient, which for the pancake model solution~(\ref{eq1_1})-(\ref{eqA4}) has the form 
\begin{equation}
[\partial v_i/\partial a_j]
= \left(\begin{array}{ccc}
-\beta_1 & 0 & 0 \\
0 & \beta_2 & 0 \\
-\omega_2 & 0 & \beta_3
\end{array}
\right).
\label{eqB2}
\end{equation}
In Fig.~\ref{fig2}(b) we compare at different times the logarithmic derivatives $\beta_{1}=-\dot{\ell}_1/\ell_1$ and $\beta_{2}=\dot{\omega}_{\max}/\omega_{\max}$ with the velocity gradients $-\partial v_1/\partial a_1$ and $\partial v_2/\partial a_2$ evaluated at the global vorticity maximum. 
One can see a very good overall agreement, with a larger deviation for $-\partial v_1/\partial a_1$ at $t = 7$. 
This deviation can be attributed to the variation (up to 20\%) of the velocity derivative within the pancake thickness.
At the final simulation time $t = 7.75$ and at the global vorticity maximum, the numerical velocity gradient is given by 
\begin{equation}
[\partial v_i/\partial a_j]_{\mathbf{a} = 0}
= \left(\begin{array}{ccc}
-0.72 &  -0.04 & -0.03 \\
-0.11 &  0.53 &  -0.09 \\
-18.42 & -0.04 &  0.19
\end{array}
\right),
\label{eqB2b}
\end{equation}
confirming that there is a single large $(3,1)$-component, $\partial v_3/\partial a_1 \approx -\omega_{\max}$. 
The diagonal components corresponding to the straining flow are in a very good agreement with the coefficients $-\beta_{1}=-0.74$, $\beta_{2}=0.53$ and $\beta_{3}=0.21$, while the remaining components, $(1,2)$, $(1,3)$, $(2,1)$, $(2,3)$ and $(3,2)$ corresponding to vanishing elements in (\ref{eqB2}) are small.

\begin{figure}
\centering
\includegraphics[width=14.0cm]{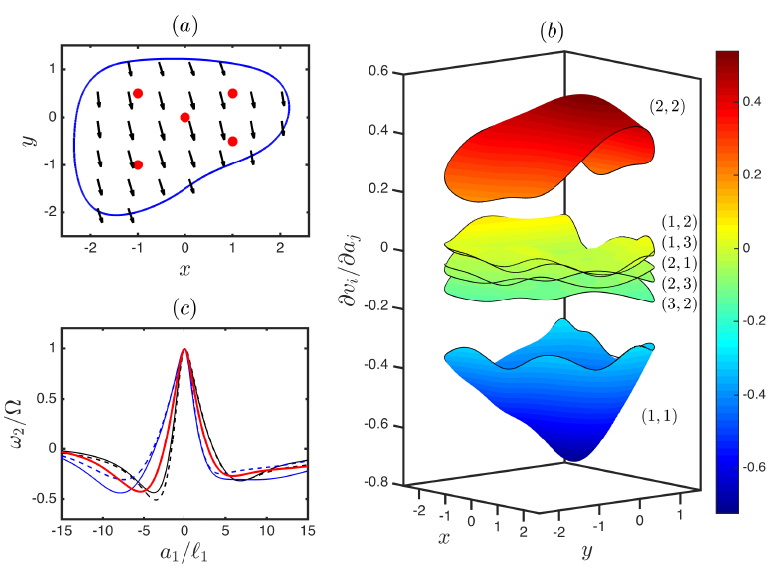}
\caption{
(a) Pancake mid-plane (vorticity above 70\% of the maximum value) in the projection to the $(x,y)$-plane. Arrows show projections of the vorticity vector, scaled by the factor of $0.025$. The coordinates are shifted to place the vorticity maximum at the origin.
(b) Velocity derivatives $\partial v_i/\partial a_j$ evaluated at different points of the pancake mid-plane; marked with $(i,j)$. 
(c) Normalized vorticity profile $\omega_{2}/\Omega$ in the direction perpendicular to the mid-plane, at five different points corresponding to red dots in the upper panel. 
The red line corresponding to the center of the pancake is the same as in Fig.~\ref{fig2}(a).
}
\label{fig3}
\end{figure}

The previous tests confirmed that the pancake model solution agrees with the flow in the vicinity of the global vorticity maximum. 
However, this model cannot describe the whole region of high vorticity, since the pancake is not completely flat, with deviations from plane $a_{1}=0$ much larger than the pancake thickness, see Fig.~\ref{fig:omega_max_dim_main}(b). 
Nevertheless, we can check if the model approximates the pancake locally, at every nearly flat pancake segment. 
For this purpose, we consider the final time and isolate the principal part of the pancake with vorticity $\omega \ge 0.7\,\omega_{\max}$; the isolated region is roughly parallel to the $(x,y)$-plane. 
Despite this region is very thin, $\ell_1 \sim 0.01$, its lateral dimensions are comparable to the size of the numerical box, as shown by the projection to the $(x,y)$-plane in Fig.~\ref{fig3}(a). 
Within this region we define the pancake mid-plane $z = z_m(x,y)$, chosen as points of maximum vorticity, $\max_z \omega$, for the given values of $x$ and $y$. 
Then, at each point $\mathbf{r}_m = (x,y,z_m(x,y))$ of the mid-plane, we define the new local coordinates $(a_1,a_2,a_3)$, using the first eigenvector of the Hessian matrix and the vorticity vector, as described above in this Section. 
Note that $(x,y)$ serve as parameters in this representation, while the coordinates $(a_1,a_2,a_3)$ explore the neighborhood of $\mathbf{r}_m$.

First, we verified numerically that, within the pancake, the vorticity vector is tangent to the mid-plane and almost unidirectional. 
This is illustrated in Fig.~\ref{fig3}(a), where projections to the $(x,y)$-plane of the vorticity vector are shown by arrows, which are scaled (multiplied by $0.025$) to fit the plot. 
At several distant points on the mid-plane, we considered the vorticity profile $\omega_{2}/\Omega$ as a function of $a_{1}/\ell_{1}$, where $\Omega = \max_z \omega$ is the vorticity at the corresponding point on the mid-plane. 
At every point, a convergence similar to Fig.~\ref{fig2}(a) is observed, showing that the vorticity profile changes with time self-similarly according to~(\ref{eqB1}); the convergence gets worse near the pancake border. 
The vorticity profile varies from point to point; this means that the function $f(\xi)$ of the model~(\ref{eq1_1}) changes along the pancake. 
The magnitude of this variation can be seen in Fig.~\ref{fig3}(c), which presents the final-time vorticity profiles $\omega_{2}/\Omega$ at five different points marked with red dots in panel (a) of the same figure. 
Note that a specific form of $f(\xi)$ is not important in the exact solution~(\ref{eq1_1})-(\ref{eqA2}). 
Thus, this function accounts for the local self-similarity only, while its variations along the pancake may be captured by a next-order correction to our model. 

The structure of the gradient~(\ref{eqB2}) is confirmed in Fig.~\ref{fig3}(b) along the whole pancake mid-plane. 
The components $(1,2)$, $(1,3)$, $(2,1)$, $(2,3)$ and $(3,2)$ are concentrated in the middle of the figure: they are about one order of magnitude smaller than the diagonal components and more than two orders of magnitude smaller than the $(3,1)$-component related to the vorticity. 
The large $(3,1)$-component does not fit to the vertical range of the figure, but it is in excellent agreement with the vorticity, $\partial v_3/\partial a_1 \approx -\omega$, with the difference below 0.6\%. 
The diagonal components $\partial v_1/\partial a_1$ (blue) and $\partial v_2/\partial a_2$ (red) vary significantly along the mid-plane. 
We do not show $\partial v_3/\partial a_3$ due to its exact relation to $(1,1)$- and $(2,2)$-components, originating from incompressibility of the flow, $\mathrm{div}\,\mathbf{v}=0$. 
With these observations we confirm that, for every nearly flat pancake segment, the flow can be approximated by the pancake model solution suggested in Section~\ref{Sec:Solution}. 

\begin{figure}
\centering
\includegraphics[width=8.4cm]{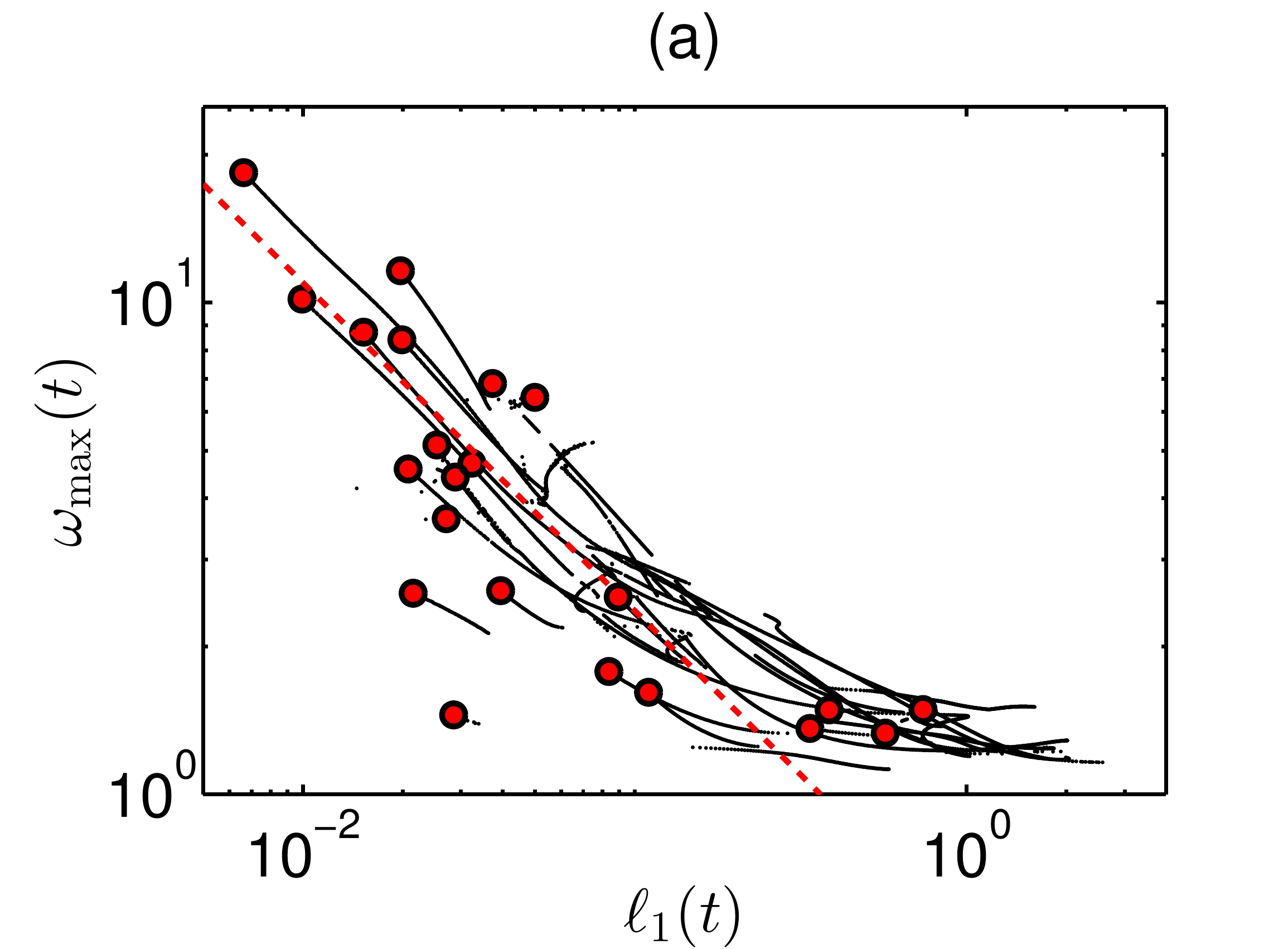}
\includegraphics[width=8.4cm]{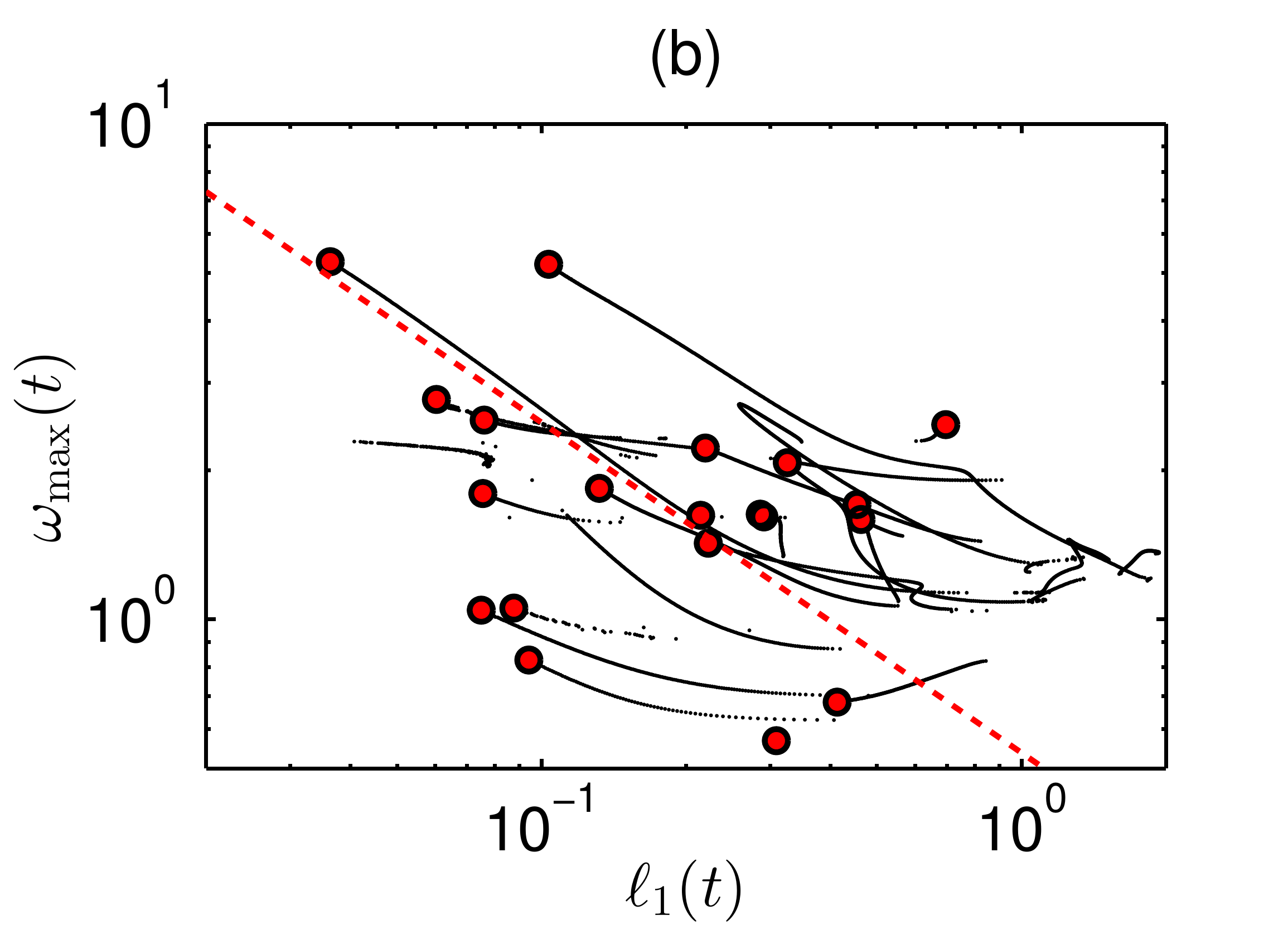}
\caption{
(a) Relation between the vorticity local maximums $\omega_{\max}(t)$ and the respective characteristic lengths $\ell_{1}(t)$ during the evolution of the pancake structures. Lines represent evolution of the maxima with increasing time, with the red dots corresponding to the final simulation time. The dashed red line indicates the power-law scaling $\omega_{\max} \propto \ell_1^{-2/3}$. 
(b) The same graph for a different simulation with a generic initial condition and the final grid $1152\times 972\times 864$. 
}
\label{fig4}
\end{figure}

As was noticed in~\cite{agafontsev2015}, at late times the pancakes develop according to the Kolmogorov-like power-law 
\begin{equation}
\omega_{\max}(t) \sim \ell_1(t)^{-\zeta},\quad \zeta \approx 2/3. 
\label{eqAA2}
\end{equation}
This tendency is clearly seen in Fig.~\ref{fig4}(a), where all other local vorticity maxima are shown and the $2/3$-slope is represented by the dashed line. 
We stress that the pancake model solution~(\ref{eq1_1})-(\ref{eqA4}) allows for an arbitrary power-law exponent $\zeta$. 
Thus, the universality of the asymptotic value $\zeta=2/3$, which corresponds to the asymmetry parameter $\sigma=1/3$ of the straining flow, goes beyond this solution. 
We think that restrictions on the power-law may come from the nonlocal interactions of the pancakes and the background flow. 

All the results presented so far related to the main pancake structure from the simulation of $I_{1}$ initial condition. 
We confirmed that several other pancake structures, associated with some of the local vorticity maxima shown in Fig.~\ref{fig4}(a), also agree with the pancake model solution. 
We performed a series of simulations in grids with $1024^{3}$ total number of nodes, starting from fully generic (large-scale) initial conditions. 
The regions of high vorticity, emerging from such initial flows, have arbitrary orientation, that does not allow to use anisotropic grids effectively; thereby, such simulations yield considerably smaller overall vorticity enhancement. 
However, these regions represent pancake-like structures developing close to the model solution~(\ref{eq1_1})--(\ref{eqA4}), and the same relation~(\ref{eqAA2}) between the vorticity maximum and pancake thickness is observed; see the example of one such simulation in Fig.~\ref{fig4}(b). 


\section{Conclusions}
\label{Sec:Conclusions}

We have studied high-vorticity regions developing in the 3D incompressible Euler equations from generic large-scale initial conditions. 
These regions represent pancake-like structures of increasing vorticity, which compress in a self-similar way in the transversal direction. 
Led by this observation, we proposed a novel \textit{exact} solution of the 3D Euler equations, which describes this behavior asymptotically. 
The proposed solution combines a shear flow aligned with an asymmetric irrotational straining flow, and is characterized by a single asymmetry parameter and an arbitrary transversal vorticity profile. 
Note that a pancake structure is not completely flat with deviations much larger than the pancake thickness. 
It is remarkable that the proposed analytical model describes locally an every nearly flat pancake segment, while the model parameters may change from one segment to another. 
The latter may be captured as next-order corrections to our pancake model, which is an interesting topic for future studies. 
In simulations, we observe exponential evolution of the vorticity maximum and pancake thickness, with the Kolmogorov-like relation between the two, $\omega_{\max}(t)\propto\ell_{1}(t)^{-2/3}$. 
This behavior is not required by the suggested model, and presumably relates to nonlocal effects. 

\vspace{5mm}
Development of the numerical code and simulations were supported by the Russian Science Foundation (grant 14-22-00174), with the latter performed at the Novosibirsk Supercomputer Center (NSU). 
Analysis of the results was done at the Data Center of IMPA (Rio de Janeiro).
D.S.A. acknowledges the support from IMPA during the visits to Brazil. 
A.A.M. was supported by the CNPq (grant 302351/2015-9) and the Program FAPERJ Pensa Rio (grant E-26/210.874/2014).


\end{document}